\documentclass[a4paper]{interact}

\usepackage{color}

\usepackage{amsmath}
\usepackage{listings}
\usepackage{diagbox}

\usepackage[caption=false]{subfig}

\usepackage[numbers,sort&compress]{natbib}
\bibpunct[, ]{[}{]}{,}{n}{,}{,}
\renewcommand\bibfont{\fontsize{10}{12}\selectfont}
\makeatletter
\def\NAT@def@citea{\def\@citea{\NAT@separator}}
\makeatother

\theoremstyle{plain}

\theoremstyle{definition}

\theoremstyle{remark}

\usepackage{braket}

\begin{document}

\articletype{ARTICLE TEMPLATE}

\title{ Rapid, Comprehensive Search of Crystalline Phases from X-ray Diffraction in Seconds via GPU-Accelerated Bayesian Variational Inference }

\author{
    \name{
        Ryo Murakami\textsuperscript{*a}, Kenji Nagata\textsuperscript{b}, Yoshitaka Matsushita\textsuperscript{a}, Masahiko Demura\textsuperscript{a}
            \thanks{
                CONTACT Ryo Murakami. Email: MURAKAMI.Ryo@nims.go.jp
            }
    }
    \affil{
        \textsuperscript{a}Research Network and Facility Services Division, National Institute for Materials Science, Tsukuba 305-0044, Japan,\\
        \textsuperscript{b}Center for Basic Research on Materials, National Institute for Materials Science, Tsukuba 305-0044, Japan,
    }
}

\maketitle

\begin{abstract}
In analysis of X-ray diffraction data, identifying the crystalline phase is important for interpreting the material. The typical method is identifying the crystalline phase from the coincidence of the main diffraction peaks. This method identifies crystalline phases by matching them as individual crystalline phases rather than as combinations of crystalline phases, in the same way as the greedy method. If multiple candidates are obtained, the researcher must subjectively select the crystalline phases. Thus, the identification results depend on the researcher's experience and knowledge of materials science. To solve this problem, we have developed a Bayesian estimation method to identify the combination of crystalline phases, taking the entire profile into account. This method estimates the Bayesian posterior probability of crystalline phase combinations by performing an approximate exhaustive search of all possible combinations. It is a method for identifying crystalline phases that takes into account all peak shapes and phase combinations. However, it takes a few hours to obtain the analysis results. The aim of this study is to develop a Bayesian method for crystalline phase identification that can provide results in seconds, which is a practical calculation time. We introduce variational sparse estimation and GPU computing. Our method is able to provide results within 10 seconds even when analysing $2^{50}$ candidate crystalline phase combinations. Furthermore, the crystalline phases identified by our method are consistent with the results of previous studies that used a high-precision algorithm.
\end{abstract}


\section{Introduction}\label{sec:introduction}
In analysis of X-ray diffraction (XRD) data, identifying constituent crystalline phases is a fundamental and critical step that affects the process of material research. A conventional identification method is to use the main diffraction peaks of each phase. For example, the Hanawalt method, which is often used when a material is supposed to consist of multiple crystalline phases, first identifies one crystalline phase whose main peaks are in positions that match those of the experimentally observed high-intensity peaks (\textit{$2\theta$} value or \textit{d} value). Next, the same peak-matching procedure is applied to the peaks that have yet to be assigned. This method repeats this procedure until all the peaks are assigned and proposes candidates for the constituent crystalline phases.

In the conventional method, if two or more crystal phases with similar diffraction angles are present, the first identification of a phase with higher-intensity peaks may preclude the successive identification of the remaining phases. This occurs because the method repeatedly matches individual phases to determine the set of phases like a greedy strategy rather than evaluate all the possible combinations of constituent phases. Moreover, because the conventional method only compares the relative intensities and diffraction angles of a limited number of high-intensity peaks, it does not consider the entire diffraction profile. Consequently, it cannot quantitatively assess the reliability of the identification results. When multiple candidate solutions are proposed, the researcher must subjectively select the appropriate solution, leading the final identification outcome to depend on the researcher's experience and expertise.

To solve the above problems, we have developed a method for estimating the combination of crystalline phases using Bayesian inference, taking the entire profile into account. This method estimates the Bayesian posterior probability of crystalline phase combinations by performing an approximate exhaustive search of all possible combinations and taking into account all peak shapes. This method can identify the crystalline phase appropriately even when there are multiple phases with similar peaks. In addition, the reliability of the identification can be quantitatively evaluated using a probability distribution. There is, however, the problem that it takes a few hours to obtain the analysis results because it uses a fitting function with a large computational cost to search for combinations of crystalline phases and profile parameters using a sampling method based on the Markov chain Monte Carlo (MCMC) method. Specifically, an analysis using $2^{50}$ candidate crystalline phase combinations took approximately 3 hours.

The aim of this study is to develop a Bayesian method for crystalline phase identification that can provide analysis results in seconds, which is a practical calculation time. We introduce variational sparse estimation and GPU computing. As a result, our method is able to provide results within 10 seconds even when analysing $2^{50}$ given candidate crystalline phase combinations.

\begin{figure}
    \centering
    \includegraphics[width=\linewidth]{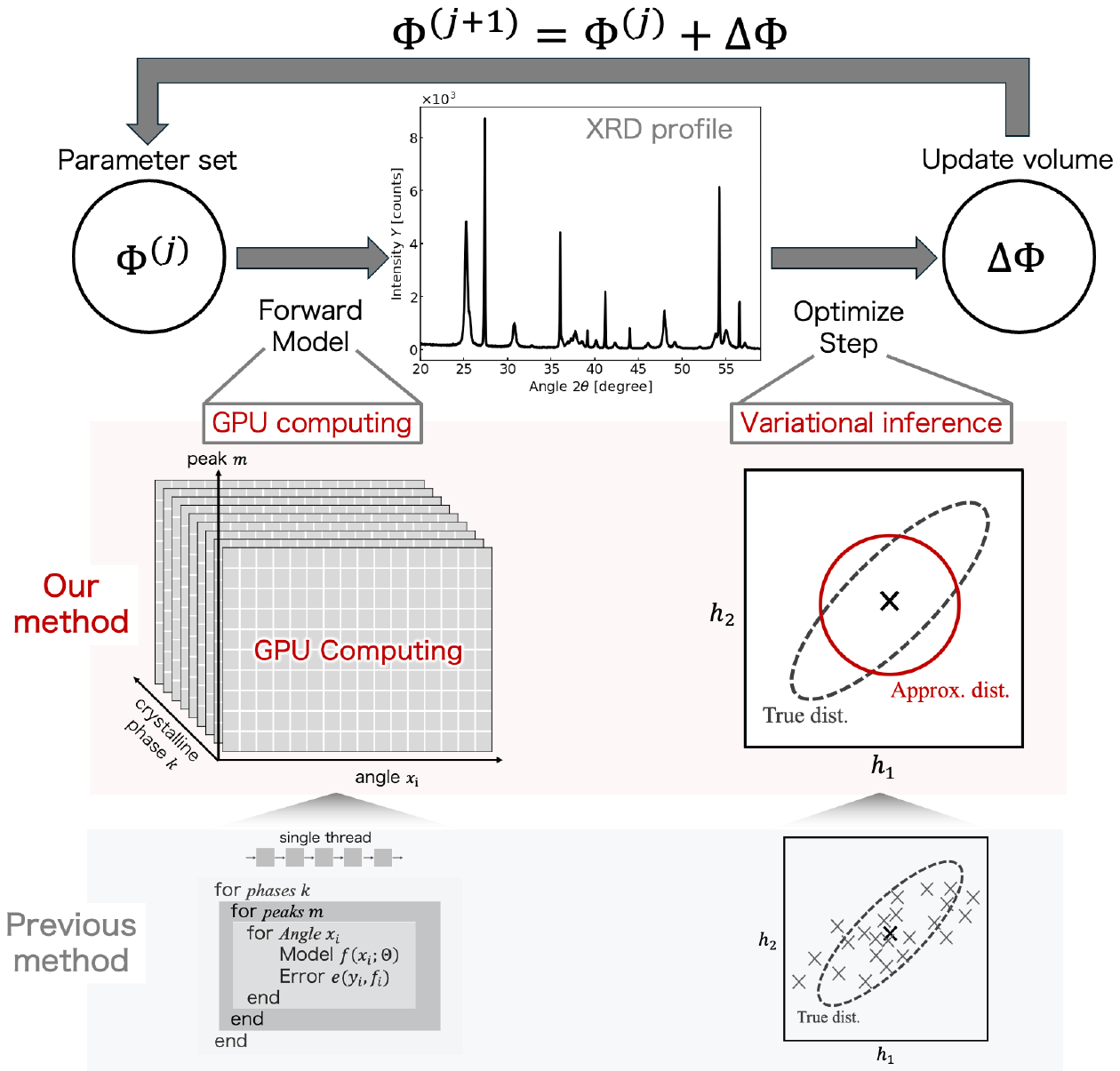}
    \caption{Overview of the accelerated strategy in our method compared with the previous method \cite{XRD_Murakami_2024}. This study tries to accelerate Bayesian crystalline identification in terms of both the algorithm and computing. The upper row figure shows the estimation flow in our method. The middle row figure shows the way to accelerate identification of crystalline phase combinations in our method. The lower row figure shows the previous method. }
    \label{fig:strategy}
\end{figure}

\section{Strategy}

\subsection{Concept for acceleration}
\quad We aim to achieve Bayesian identification of a crystalline phase combination in seconds by taking into account the overall shape of the profile. Figure \ref{fig:strategy} shows the overview of the accelerated strategy in our method compared with the previous method \cite{XRD_Murakami_2024}. This study tries to accelerate Bayesian crystalline identification in terms of both the algorithm and computing.

For the algorithm, we introduce variational inference instead of the MCMC method to accelerate crystalline phase identification using Bayesian estimation. Variational inference can reduce the computational cost and the time required for convergence compared with the MCMC method because it solves an inference problem as an optimization problem for which the gradient method can be used. However, the model of the previous method \cite{XRD_Murakami_2024} introduces discrete variables for crystalline phase identification. Because of these discrete variables, gradient-based optimization is not efficient with this model as is. To solve the problem, we apply continuous relaxation like L1 regularization to the discrete variables. In this study, sparsity is introduced into the intensity parameters using properties of the gamma distribution. Our method differs from previous works by using a continuous relaxation of the model, so it is possible to use acceleration algorithms. Our method uses stochastic variational inference (SVI) \cite{svi_Matthew} to estimate the Bayesian posterior distribution, which is faster than the MCMC method used in previous work. 

For computing, our method uses a GPU to accelerate crystalline phase identification. The rate-limiting factor in crystalline identification, which takes into account the profile shape, is calculating the profile function that generates the diffraction peak. This study focuses on the fact that XRD profile calculations can be performed independently and in parallel for parameter sets and inputs. Here, the inputs are the diffraction angle and the information on candidate crystalline phases. We can implement a function that generates peaks in the single program multiple data (SPMD) architecture \cite{spmd1983}, which allows multiple processors to work together to execute a program, enabling parallel processing to obtain results more quickly. Therefore, GPU computing enables us to generate the profile function for each crystalline phase faster than CPU computing. GPU computing is highly effective for generating XRD profiles, which involve a huge number of combinations of parameters and inputs.

\subsection{Prior distribution design for SVI}
\quad In this study, we redesign the prior distribution to introduce SVI efficiently. Our method identifies the crystalline phases using a prior distribution with sparsity for continuous variables instead of introducing discrete variables as in a previous study. The intensity parameter $h_k$ can only be positive, that is, $h_k \in \mathbb{R}^{+}_{0}$. The prior distribution of the intensity parameter is set as the gamma distribution, which is a general probability distribution with a positive value range. The gamma distribution $\mathrm{Gam}(\cdot)$ is described by the shape parameter $\alpha$ and scale parameter $\lambda$ as follows:
\begin{eqnarray}
    h_k \sim \mathrm{Gam}(h_k; \alpha, \lambda) = \frac{\lambda^{\alpha}}{\Gamma{(\alpha)}}h_k^{\alpha-1}\exp{(-\lambda h_k)}.
\end{eqnarray}
The expected value and variance in a gamma distribution are
\begin{eqnarray}
    \mathbb{E}_{\mathrm{Gam}(h_k; \alpha, \lambda)}[h_k] = \frac{\alpha}{\lambda}, \,\,\, \mathbb{V}_{\mathrm{Gam}(h_k; \alpha, \lambda)}[h_k] = \frac{\alpha}{\lambda^2}.
\end{eqnarray}
Here, we apply sparse modelling to the intensity parameter $h_k$ based on the assumption that only a few of the candidate combination phases are included in the measurement data. That is, we set a prior distribution with sparse intensity $h_k$. We set a gamma distribution with the shape parameter $\alpha=1$ (which is the exponential distribution) as the prior distribution to achieve sparsity:
\begin{eqnarray}
    h_k \sim \mathrm{Gam}(h_k;\alpha=1, \lambda) = \lambda \exp{(-\lambda h_k)}.
\end{eqnarray}
This prior distribution setting is equivalent to the Laplace distribution (L1 regularization) with a positive range. Therefore, the estimated intensity parameters have sparsity. The hyperparameter $\lambda$ corresponds to the intensity scale from the definition of the gamma distribution. Therefore, we set the maximum measured signal, $\eta = \mathrm{max}(Y-B)$, to be the expected value. In other words, because the shape parameter is set as $\alpha=1$, we set the scale parameter as $\lambda = \eta^{-1}$.

\begin{figure}
    \centering
    \includegraphics[width=\linewidth]{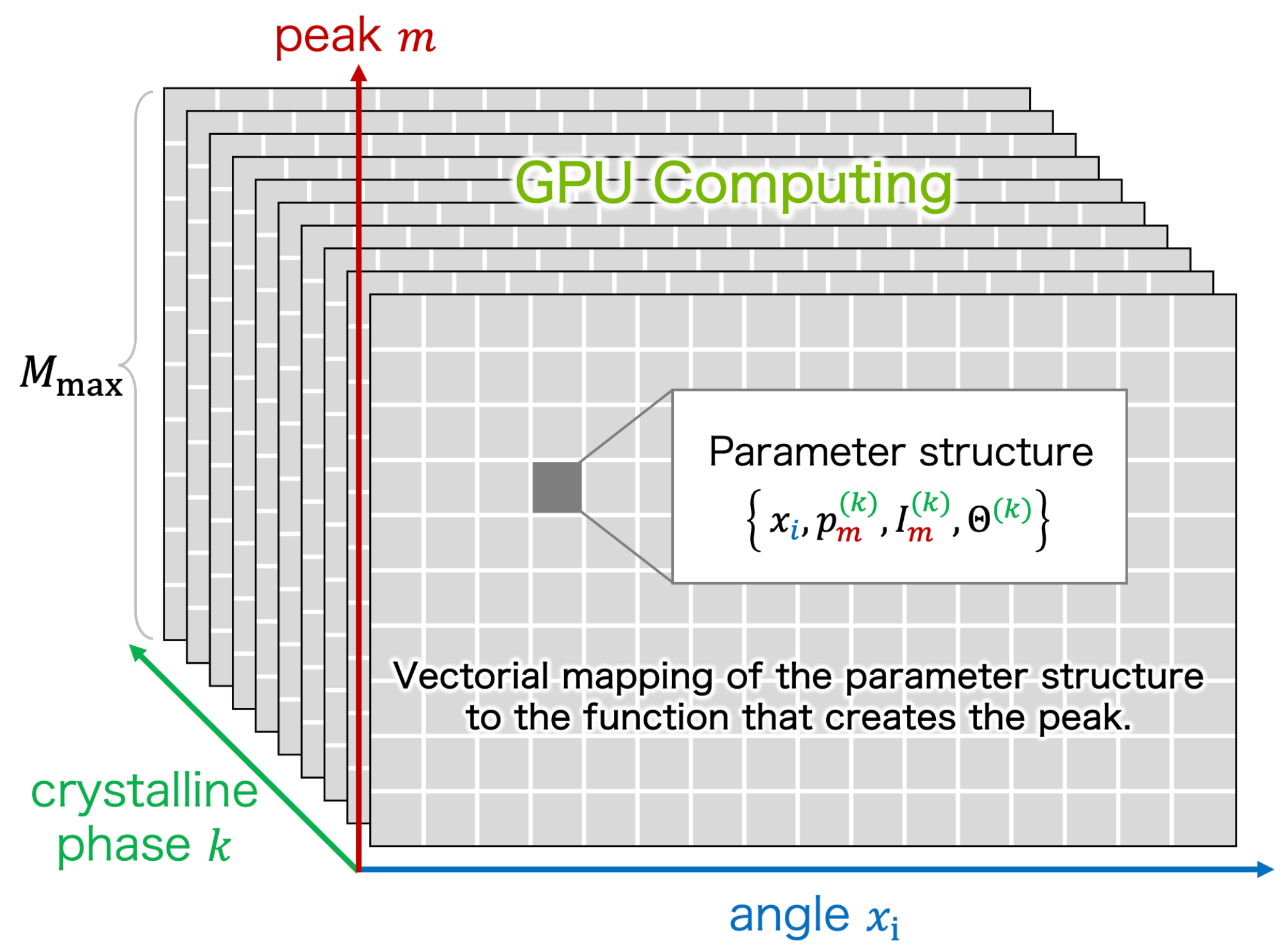}
    \caption{Tensor representation of the parameter structure for calculating peak profiles on a GPU. It is constructed by mapping functions to tensors.}
    \label{fig:tensor}
\end{figure}

\subsection{GPU computing to generate an XRD profile}
\quad For crystalline phase identification that takes into account the profile shape, the bottleneck in calculation time is due to calculating the function that generates the peaks. In XRD data analysis, the number of candidate crystalline phase factors $\mathcal{F}_k$ and number of peaks $M_k$ are huge. Therefore, to calculate the fitting model, it is necessary to calculate the triple loop $\sum^{K}_{k=1}{ \sum_{m=1}^{M_k}{ \sum^{N}_{i=1}{ \mathrm{Forword}(x_i, p^{(k)}_{m}, I^{(k)}_{m}; \Phi^{(k)}) } } }$ to get the forward model. The inputs ($x_i$, $p^{(k)}_{m}$ and $I^{(k)}_{m}$) are given static values, and the parameter set $\Phi^{(k)}$ is updated for each optimization step.

This study focused on the fact that XRD profile calculations can be performed independently and in parallel for parameter set $\Phi^{(k)}$ and inputs ($x_i$, $p^{(k)}_{m}$ and $I^{(k)}_{m}$). A peak is generated using SPMD architecture, which it is suitable for vector mapping and calculations in GPU computing. Figure \ref{fig:tensor} shows the tensor representation of the parameter structure for generating peaks for vector mapping. As shown in the figure, we apply vector mapping of the parameter structure into functions that generate a peak, and the functions form the tensor. Furthermore, vectorization makes it possible to use GPU computing. In our method, the peak-generating function is vectorized on the GPU to enable high-speed calculation.

We used the JAX library \cite{jax2018github}, which is developed by Google, to implement vectorized mapping of the function that generates the peaks. Vectorizing functions can be easily implemented using the $vmap$ method in JAX. We can map to tensor space by applying $vmap$ multiple times. JAX is GPU-compatible, so it can be run on a GPU without any changes to the programming code.

If the number of peaks $M_k$ differs for each crystalline phase, it is not possible to express it as a tensor. For this reason, it is necessary to introduce a dummy variable so that $M_k$ becomes $M_{max}$ in the implementation. $M_{max}$ is the maximum value of $M_k$. The dummy variable should be set so that it does not affect the profile.

\section{Method}

\subsection{XRD data for verification}
\quad The measurement sample was a mixture of multiple types of $\rm TiO_2$: anatase, brookite, and rutile. The mixture ratios were equal (1/1/1 wt. \%). We prepared measurement samples such that the crystalline phases were homogeneous. Consequently, we measured the XRD data using monochromatic X-rays of $\rm{Cu} \ \rm{K}_{\alpha1}$. A non-reflecting plate cut from a specific orientation of a single crystal of silicon was used as the sample plate. The diffraction angles ($2\theta$) were in the range 10--60 [$^{\circ}$], with the values of $2\theta$ corresponding to $\bm x = (10.00, 10.02, 10.04, ..., 60.00)^{\top}$. 

Our method tried to estimate the crystalline phase in measurement samples from 50 candidates. Candidate crystalline phases were obtained from AtomWork-Adv \cite{AtomWork} as described in the literature \cite{XRD_Murakami_2024}.

\subsection{Algorithm for estimating an approximate posterior distribution}
\quad We assume that the observed data $\{(Y, X)\}=\{(x_i,y_i)\}_{i=0}^{N}$ are stochastically distributed owing to statistical noise in the measurement. Our aim is to estimate the posterior distribution $p(\Phi \mid Y)$ of the parameter set $\Phi$. First, we consider the joint distribution $p(Y, \Phi)$, which can be expanded to $p(Y, \Phi)=p(\Phi \mid Y)p(Y)$. Using Bayes' theorem to swap the orders of $Y$ and $\Phi$, we can expand $p(Y, \Phi)=p(Y \mid \Phi)p(\Phi)$. Hence, the posterior distribution $p(\Phi \mid Y)$ is expressed as
\begin{eqnarray}
    p(\Phi \mid Y) = \frac{p(Y \mid \Phi)p(\Phi)}{p(Y)},
\end{eqnarray}
where $p(\Phi \mid Y)$ and $p(\Phi)$ are the posterior and prior distributions, respectively, in the Bayesian inference; $p(Y \mid \Phi)$ is the conditional probability of $Y$ under given the model parameter set $\Phi$, which is a probability distribution explained by statistical noise. We call the probability $p(Y \mid \Phi)$ the likelihood. The normalized constant $p(Y)$ is expressed as 
\begin{align}
    p(Y) = \int{p(Y, \Phi)}\mathrm{d}\Phi = \int{p(Y \mid \Phi)p(\Phi)}\mathrm{d}\Phi.
\end{align}
This is called the marginal likelihood, which is an important measure of how much the model explains the observation data.

We introduce variational inference to rapidly obtain this posterior distribution. Variational inference is a method for approximating the posterior distribution $p(\Phi \mid Y)$ using a specific family of distributions $q(\Phi)$ that are easy to deal with (e.g. the Gaussian distribution): $p(\Phi \mid Y) \approx q(\Phi)$. Assuming a specific distribution and parameter independence makes it possible to estimate the posterior distribution in the same way as in a gradient-based optimization problem.

The marginal log-likelihood $\ln{p(Y)}$, that is, the model evidence, can be decomposed into two functional terms:
\begin{align}
    \ln{ p(Y) } = \mathcal{L}[q] + \mathrm{KL}[q||p] \geq \mathcal{L}[q],
\end{align}
where KL represents the Kullback-Leibler divergence, and we set $q=q(\Phi)$ and $p=p(\Phi \mid Y)$. These functional terms denote
\begin{align}
    \mathcal{L}[q] := \int{q(\Phi) \ln{ \frac{p(Y|\Phi)P(\Phi)}{q(\Phi)} } \mathrm{d}\Phi},\\
    \mathrm{KL}[q||p] := - \int{q(\Phi) \ln{ \frac{p(\Phi \mid Y)}{q(\Phi)}} \mathrm{d}\Phi}.
\end{align}
According to the definition of the KL divergence, it is impossible for $\mathrm{KL}[q||p]$ to be negative, that is, $\mathrm{KL}[q||p] \geq 0$. Therefore, $\mathcal{L}[q]$ becomes the lower bound of model evidence $\ln{p(Y)}$. That is why $\mathcal{L}[q]$ is generally called the evidence lower bound (ELBO) \cite{kingma2013auto}. We can estimate the approximate posterior distribution $\hat{q}$ by maximizing the ELBO $\mathcal{L}[q]$:
\begin{align}
    \hat{q} = \underset{q} {\operatorname{argmax}} \: \mathcal{L}[q].
\end{align}
To implement the variational Bayesian inference, we consider restricting the distribution class of $q(\Phi)$. In this study, we assume that $q(\Phi)$ can be expressed as the mean field approximation $\prod_{\phi \in \Phi}{q(\phi)}$. Moreover, we assume that each posterior distribution can be approximated by a Gaussian distribution $\mathcal{N}(\phi; \mu, \sigma^{2})$:
\begin{align}
    q(\Phi) \approx \prod_{\phi \in \Phi}{q(\phi)} \approx \prod_{\phi \in \Phi}{\mathcal{N}(\phi; \mu=\mu_{\phi}, \sigma^{2}=\sigma^{2}_{\phi})},
\end{align}
where hyperparameters $\mu$ and $\sigma^2$ are the mean and standard deviation of the Gaussian distribution, respectively. We estimate these hyperparameters using variational inference based on the ELBO to obtain the approximate posterior distribution.

\section{Results and discussion}
\begin{figure}
    \centering
    \includegraphics[width=\linewidth]{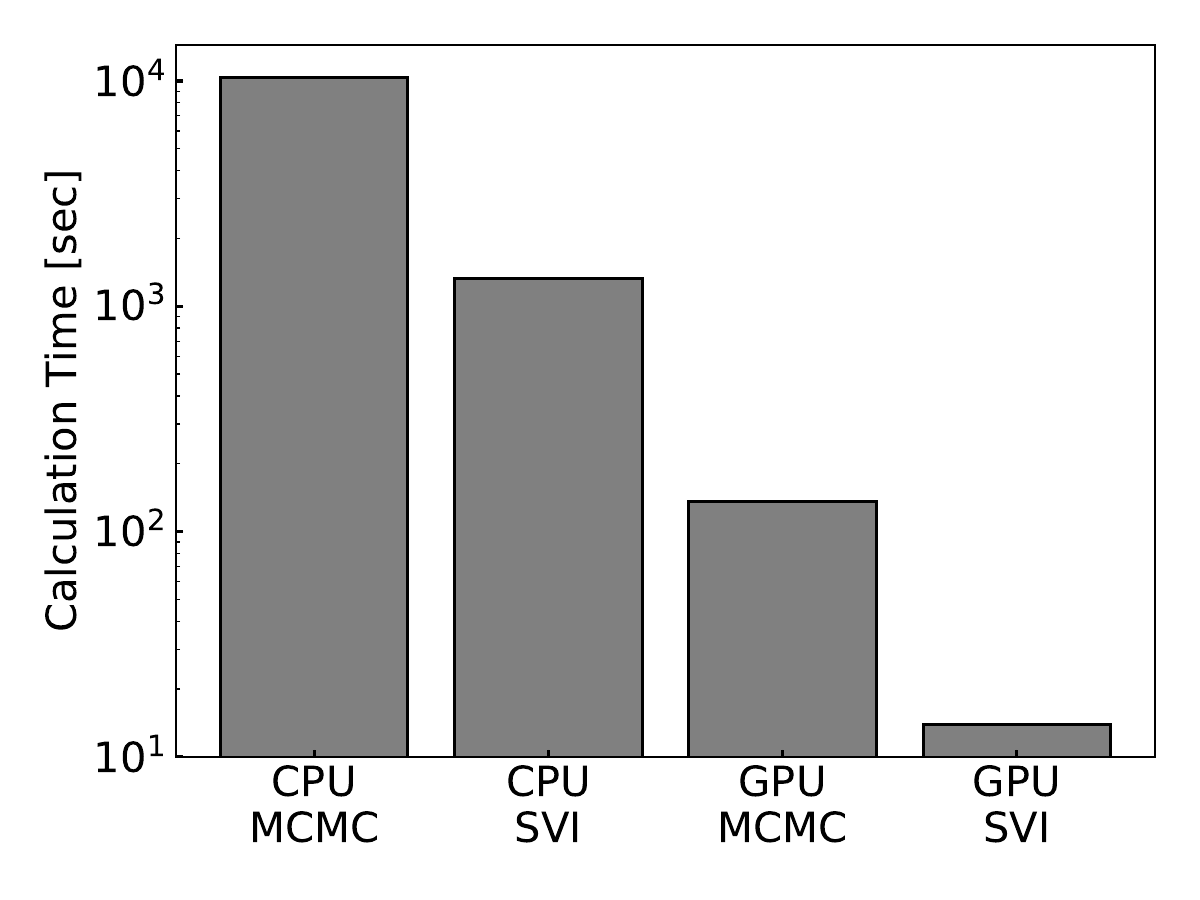}
    \caption{Execution times for each algorithm and implementation. }
    \label{fig:speed}
\end{figure}

\quad First, we compare the execution times of different algorithms and different computations. There was no significant difference between the computational efficiencies of our modelling and the previous modelling in the MCMC method. Therefore, to make comparisons easier, the model was unified with the prior distribution we designed. In this study, we conducted computational experiments on a total of four combinations using either SVI or the MCMC method for the algorithm and either a CPU or GPU for the computing. In the MCMC method, we performed 1000 sampling steps and 1000 burn-in steps. For the SVI, we performed 1000 optimization steps. Figure \ref{fig:speed} and Table \ref{table:times} show the execution times for each algorithm and computation. In the figure, the y-axis denotes calculation time [second] for the log-scale. The results for identifying the crystalline phase are consistent across all methods. The method using the GPU and SVI was the fastest of all these methods, providing the results in 7.2 seconds for this case. We have achieved a practical time for Bayesian estimation of a crystalline phase combination that takes the overall profile shape into account.

\begin{table}
    \centering
    \caption{Execution times [seconds] of different algorithms and computing implementations.}
    \label{table:times}
    \scalebox{1.2}[1.2]{
    \begin{tabular}{c|cc}
        \hline
        \diagbox{Algorithm}{Computing} & CPU & GPU \\
        \hline
        MCMC method  & 10371.5 sec & 1326.9 sec \\
        SVI metohod  & 260.8 sec   & 7.2 sec  \\
        \hline
    \end{tabular}
    }
\end{table}

\begin{figure}
    \centering
    \includegraphics[width=\linewidth]{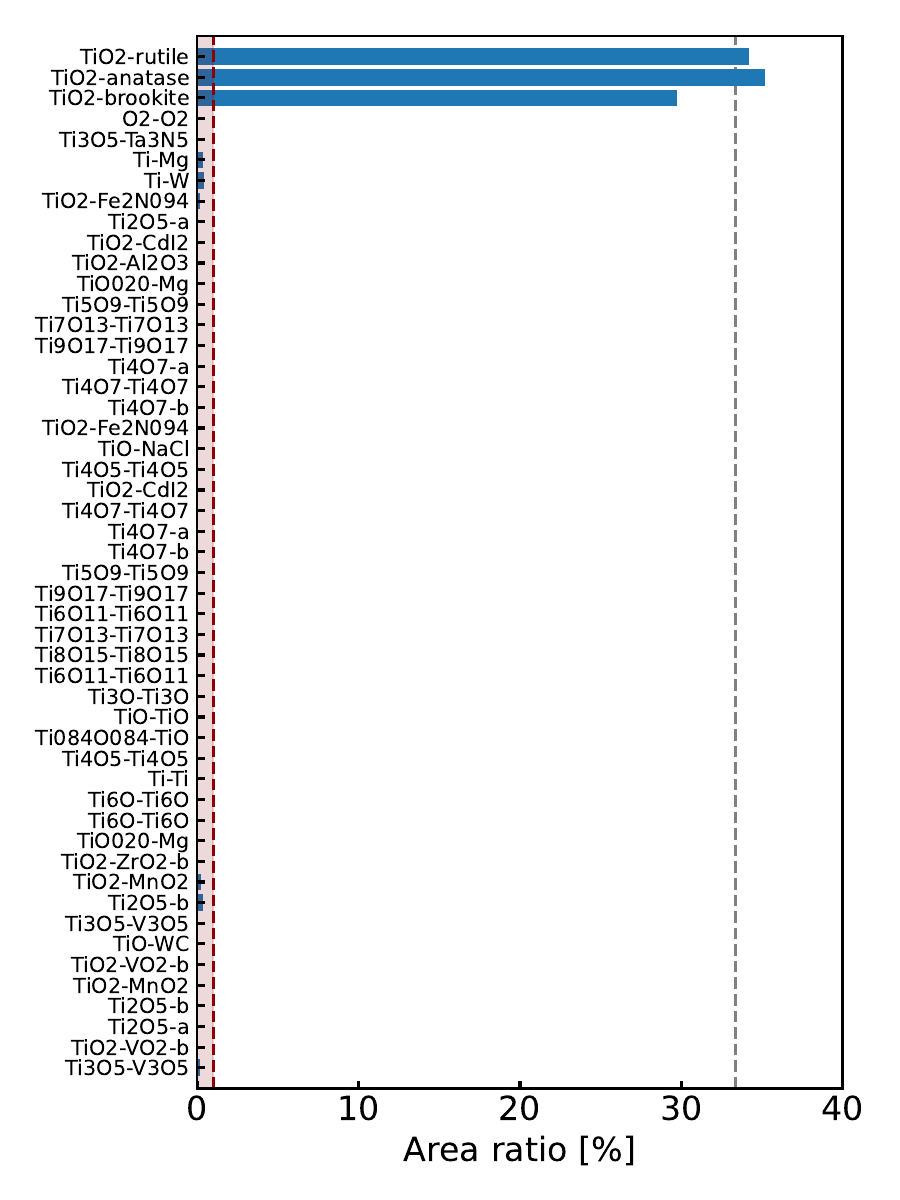}
    \caption{ Area ratios [\%] estimated by our method for each crystalline phase. }
    \label{fig:selection}
\end{figure}

\begin{figure}
    \centering
    \includegraphics[width=\linewidth]{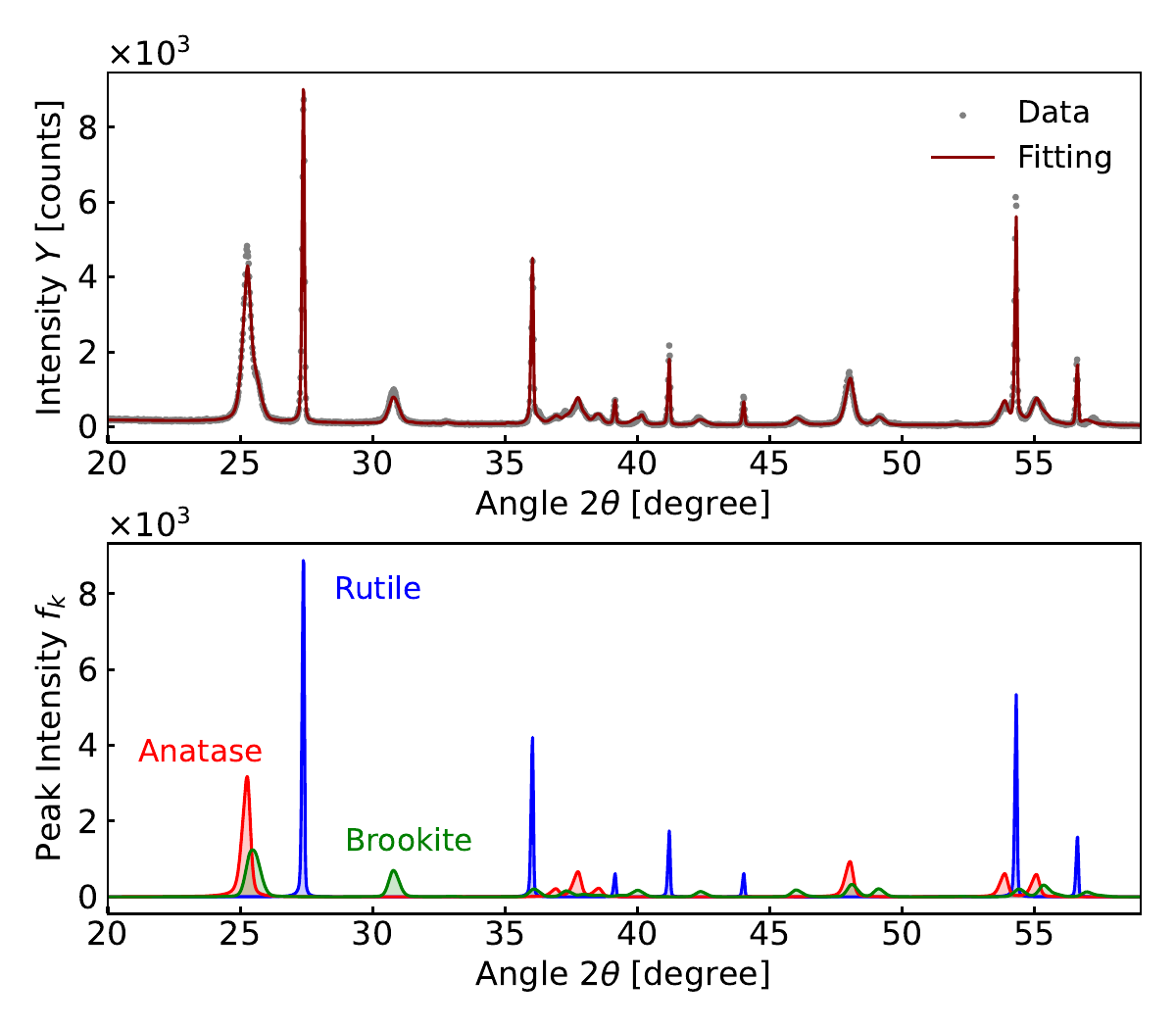}
    \caption{ Results of profile analysis for measured XRD data using our method. (a) Fitting results via the profile function for the measured XRD data. (b) Peak components of the three crystalline phases of $\rm TiO_2$: anatase, brookite, and rutile. }
    \label{fig:fitting}
\end{figure}

We show the selection result for our method using shrinkage estimation. Figure \ref{fig:selection} shows the resulting area ratios [\%] estimated by our method for each crystalline phase. The x- and y-axes denote the area ratio and crystalline phases, respectively. The grey dashed line is the prepared mixing ratio of the measured sample. The red dashed line is the threshold for an area fraction of 1 [\%] or less. This figure confirms that the intensity not included in the measurement sample was reduced to zero because our method used a distribution with sparsity as the prior distribution of the intensity parameters. Our method allows identification of three true crystalline phases of $\rm TiO_2$: anatase, brookite, and rutile. Furthermore, the estimated ratio is close to the prepared mixing ratio.

Figure \ref{fig:fitting}(a) shows the fitting results via the profile function for the measured XRD data. In Figure \ref{fig:fitting}(a), the black and red lines indicate the measured XRD data and the fitting profile functions, respectively. Figure \ref{fig:fitting}(b) shows the peak components of the three crystalline phases anatase, brookite, and rutile, indicated by the red, green, and blue lines, respectively. The estimated profile function facilitates a good fit of the XRD data. Even though we analysed 50 candidate crystalline phases, the analysis succeeded in providing results in 7.2 seconds.

\begin{figure}
    \centering
    \includegraphics[width=\linewidth]{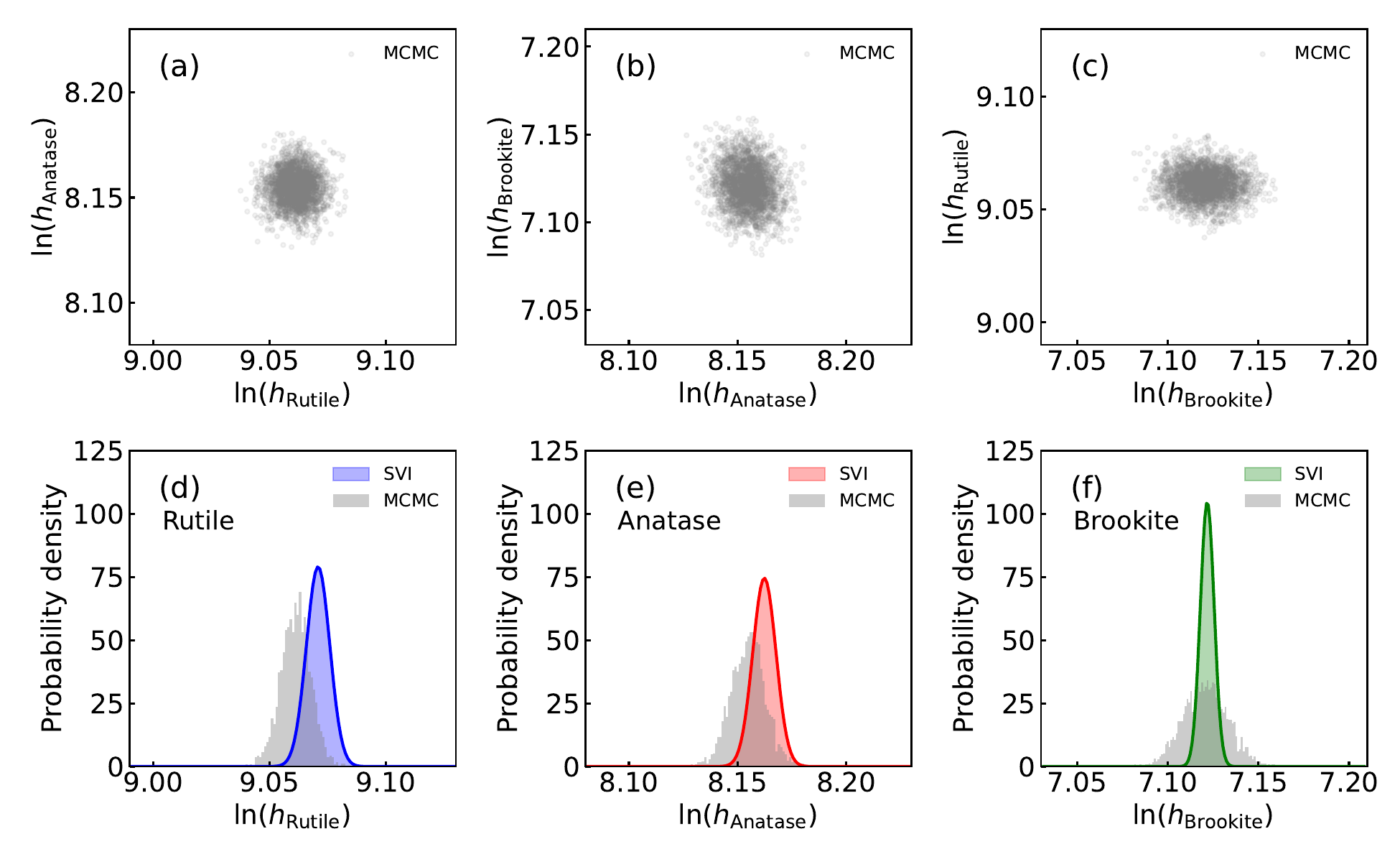}
    \caption{Posterior distributions of intensity parameters. (a)-(c) Two-paired posterior distribution of intensity parameters. (d)-(f) Comparisons of posterior distributions obtained by the MCMC method and SVI. }
    \label{fig:compare_post_h}
\end{figure}

We used SVI instead of the MCMC method to identify crystalline phases quickly. In contrast to the MCMC method, SVI estimates the posterior distribution assuming that there is no correlation between the parameters, which is the mean field approximation. While SVI can achieve high-speed estimation, the mean field approximation is not always appropriate depending on the model and data. We focus on the intensity parameters of anatase, brookite, and rutile and determine whether there is any correlation in the posterior distribution obtained by the MCMC method. Furthermore, we compare the posterior distributions obtained using the MCMC and SVI methods. Figure \ref{fig:compare_post_h}(a)-(c) shows the two-paired posterior distribution of intensity, which is the primary parameter. Figure \ref{fig:compare_post_h}(d)-(f) shows comparisons of posterior distributions obtained by the MCMC method and SVI. Panels (a)-(c) do not show an effective correlation in intensity parameters. This suggests that the mean field approximation is a reasonable assumption for the intensity parameters. In panels (d)-(f), each MAP solution is similar. However, there are differences in the width of the estimated posterior distribution. SVI underestimates the width of the posterior distribution of the strength parameter of brookite. This might be due to the poor crystallinity of brookite and its small integral intensity compared with other crystalline phases.

\section{Conclusion}
\quad We aimed to develop a Bayesian method that can identify crystalline phases in seconds using variational sparse inference and GPU computing. This method succeeded in providing results in 7.2 seconds even though we analysed $2^{50}$ candidate crystalline phase combinations. Furthermore, the crystalline phases identified by our method were consistent with the precise calculations of the MCMC method. We have achieved identification in seconds with Bayesian estimation of crystalline phases that takes the overall profile shape into account.

\section*{Acknowledgement}
This work was supported by GteX Program Japan Grant Number JPMJGX23S6. We would like to thank Dr. Hayaru Shouno (The University of Electro-Communications) and Dr. Hideki Yoshikawa (NIMS) for useful discussions.

\bibliographystyle{unsrt} 
\bibliography{main} %

\appendix

\section{Model}\label{sec:model}
\subsection{Problem setting}
\quad The purpose is to estimate the profile parameters and crystalline phase structures in the measured sample, considering the measured XRD data $\mathcal{D}=\{(x_i, y_i)\}_{i=1}^{N}$ and the candidate crystal structure $\mathcal{F}$. Here, $x_i \in (0, 180)$ and $y_i \in \mathbb{N}$ denote the diffraction angle $2\theta$ [$^{\circ}$] and diffraction intensity [counts], respectively.

The candidate crystal structure factor set $\mathcal{F}$ is expressed as
\begin{eqnarray}
    \mathcal{F} &=& \{\mathcal{F}_{k} \mid k \in \{1, 2, ..., K\}\},\\
    \text{where} \:\: \mathcal{F}_{k} &=& \{(p^{(k)}_{m}, I^{(k)}_{m}) \mid m \in \{1, 2, ..., M_k\}\} \subset \mathcal{F},
\end{eqnarray}
where $K\in\mathbb{N}$ is the number of candidate crystal structures, and $\mathcal{F}_k$ is the $k$-th crystalline structure factor. The elements of the crystal structure factor $p^{(k)}_{m} \in (0, 180)$ and $I^{(k)}_{m} \in [0, 1]$ are the diffraction angle (peak position) [$^{\circ}$] and relative intensity of the $m$-th diffraction peak in $\mathcal{F}_k$ for a crystalline structure $k$. The symbol $M_{k}\in\mathbb{N}$ denotes the number of peaks in $\mathcal{F}_k$. In this study, the candidate crystalline structure factor set $\mathcal{F}$ is provided.

\subsection{Profile function}
\quad XRD data can be represented by a profile function $f_{\mathcal{F}}(x_i;\Phi):\mathbb{R} \rightarrow \mathbb{R}^{+}_{0}$, which is a linear sum of the signal spectrum $S_{\mathcal{F}}(x_i;\Phi)$ and the background $B(x_i;\Phi)$:
\begin{eqnarray}
        y_i &\approx& f_{\mathcal{F}}(x_i;\Phi),\\
        &=& S_{\mathcal{F}}(x_i;\Phi) + B(x_i),
\end{eqnarray}
where $(x_i, y_i)$ denote the measured data points, the function $S_{\mathcal{F}}(x_i;\Phi)$ denotes the signal spectrum based on the candidate crystal structures $\mathcal{F}$, and the function $B(x_i)$ denotes the background.

The signal spectrum $S_{\mathcal{F}}(x_i;\Phi)$ is expressed as a linear sum of the profile function (peaks) $C_{\mathcal{F}_{k}}(x_i;\Phi^{(k)}):\mathbb{R} \rightarrow \mathbb{R}^{+}_{0}$ in a crystal structure $\mathcal{F}_{k}$ among several candidates \cite{1990ProfileFunction}:
\begin{eqnarray}\label{eq:xrd:signal}
    S_{\mathcal{F}}(x_i;\Phi) &=& \sum_{k=1}^{K}{
        h_{k} C_{\mathcal{F}_{k}}(x_i;\Phi^{(k)})
    }.
\end{eqnarray}
We estimated the background $B(x_n)$ using pybaselines \cite{pybaselines} before peak extraction. The pybaselines can perform mostly model-free background estimation through the iterative least squares method.

In this study, we use SVI to estimate the posterior distribution $P(\Phi \mid Y)$. The settings for the prior distribution $P(\Phi)$ are described in Appendix \ref{sec:configuration}.

The profile function $C_{\mathcal{F}_{k}}(x_i;\Phi^{(k)})$ of candidate crystal structure $k$ is defined as 
\begin{eqnarray}
    C_{\mathcal{F}_{k}}(x_i;\Phi^{(k)}) &=& \sum_{m=1}^{M_{k}}{
        I_{m}^{(k)}
        V \left (
            x_i;
            \rho_{mk},
            w_k,
            r_k
        \right )
    },\\
    &=& \sum_{m=1}^{M_{k}}{
        I_{m}^{(k)} \{(1-r_k) G(x_i;\rho_{mk}, w_k) + r_k L(x_i;\rho_{mk}, w_k)\}},\\
    \text{where} \:\: \rho_{mk} &=& p_{m}^{(k)}+\mu_k,
\end{eqnarray}
where $\mu_k \in \mathbb{R}$ and $r_k \in [0, 1]$ are the peak shift and Gauss-Lorentz ratio at the peak of crystal structure $k$, respectively; $\rho_{mk} \in \mathbb{R}$ is the peak position of the peak function; $V(x_i):\mathbb{R} \rightarrow \mathbb{R}^{+}_{0}$ is a pseudo-Voigt function \cite{1974PseudoVoigt};  $G(x_i):\mathbb{R} \rightarrow \mathbb{R}^{+}_{0}$ and $L(x_i):\mathbb{R} \rightarrow \mathbb{R}^{+}_{0}$ are Gaussian and Lorentz functions, respectively.
\begin{eqnarray}
    \text{where} \:\: A(x_i;\alpha_{k}) &=& \left\{
        \begin{array}{ll}
        \alpha_{k} & (x_{i} \geq \rho_{k})\\
        1 & (x_{i} < \rho_{k}),
        \end{array}
        \right.\\
        &=& \operatorname{sign}(x_{i}-\rho_{k})\frac{\alpha_{k} - 1}{2} + \frac{\alpha_{k} + 1}{2},
\end{eqnarray}
Function $A(x_i;\alpha_{k}):\mathbb{R} \rightarrow \mathbb{R}$ expresses the peak asymmetry, and $\alpha_{k} \in \mathbb{R}^{+}$ is the asymmetry parameter for the peak function. Furthermore, $\mathrm{sign}(\cdot):\mathbb{R} \rightarrow \{-1, 1\}$ is the sign function, and $\sec(x)$ is the trigonometric function $\sec(x)=1/\cos(x)$.

To derive $p(Y \mid \Phi)$, we consider the observation process for $\{(x_i, y_i)\}$ at the observation data points. Assuming that the observed data are independent of each other, the conditional probability of the observed data $\{(Y, X)\}$ can be expressed as
\begin{eqnarray}
    P(Y \mid \Phi) = \prod_{i=0}^{N}{p(y_i \mid \Phi)}.
\end{eqnarray}
Because XRD spectra are count data, the conditional probability $p(y_i \mid \Phi)$ of the intensity $y_i$ for the diffraction angle $x_i$ follows a Poisson distribution $\mathrm{Poi}(y_i \mid f_{\mathcal{F}}(x_i;\Phi))$:
\begin{eqnarray}
    p(y_i \mid \Phi) &=& \mathrm{Poi}(y_i \mid f_{\mathcal{F}}(x_i;\Phi))\\
    &=& \frac{f_{\mathcal{F}}(x_i;\Phi)^{y_i}\exp{(-f_{\mathcal{F}}(x_i;\Phi))}}{y_i!}.
\end{eqnarray}
The negative log-likelihood function $-\ln{P(Y \mid \Phi)}$ is expressed as
\begin{eqnarray}
    -\ln{p(Y \mid \Phi)} &=& - \sum_{i=1}^{N}{\ln{p(y_i \mid \Phi)}},\\
                    &=& - \sum_{i=1}^{N}{\ln{\mathrm{Poi}(y_i \mid f_{\mathcal{F}}(x_i;\Phi))}},\\
                    &=& - \sum_{i=1}^{N}{
        \left\{ y_{i}\ln{f_{\mathcal{F}}(x_i;\Phi)} - f_{\mathcal{F}}(x_i;\Phi) - \ln{y_i!} \right\}
    }.
\end{eqnarray}

\section{Configuration}\label{sec:configuration}

\subsection{Calculator specification}
The calculator was an Intel Xeon(R) Platinum 8280 with a 2.70 GHz CPU (112 threads) and a Tesla V100S-PCIE-32GB GPU.

\subsection{Configuration of prior distribution}
\quad We set the prior distribution over the parameter set $\Phi$ of the profile function as follows:
\begin{eqnarray*}
    \text{Profile pattern shift:}\:\:\:\:           \mu_k &\sim& \mathrm{Norm}(\mu_N=0.00, \sigma_N=0.03),\\
    \text{Peak asymmetry:}\:\:\:\:          \alpha_k &\sim& \mathrm{Norm}(\mu_N=1.00, \sigma_N=0.20),\\
    \text{Lorentz-Gaussian ratio:}\:\:\:\:  r_k &\sim& \mathrm{Uni}(u_U=0.00, l_U=1.00),\\
    \text{Peak width parameter:}\:\:\:\:    w_k &\sim& \mathrm{Gam}(\alpha_G=3.00, \lambda_G=100.00),
\end{eqnarray*}
where the probability distribution $\mathrm{Gam}(k_G, \theta_G)$ is the gamma distribution and $\alpha_G \in \mathbb{R}^{+}$ and $\lambda_G \in \mathbb{R}^{+}$ are the shape and scale parameters, respectively. The probability distribution $\mathrm{Norm}(\mu_N, \sigma_N)$ is a normal distribution, and $\mu_N \in \mathbb{R}$ and $\sigma_N \in \mathbb{R}^{+}$ are the mean and standard deviation, respectively. The probability distribution $\mathrm{Uni}(u_U, l_U)$ is a uniform distribution, with $u_U \in \mathbb{R}$ and $l_U \in \mathbb{R}$ being the maximum and minimum values, respectively.

\end{document}